\documentstyle[preprint,subeqn,eqsecnum,aps,epsfig,floats]{revtex}

\begin{document}                                                 

\newcommand{\be}{\begin{equation}}
\newcommand{\ee}{\end{equation}}
\newcommand{\ba}{\begin{eqnarray}}
\newcommand{\ea}{\end{eqnarray}}
\newcommand{\bc}{\begin{center}}
\newcommand{\ec}{\end{center}}
\newcommand{\vs}{\vspace*{3mm}}
\newcommand{\dis}{\displaystyle}
\newcommand{\bay}{\begin{array}{rcl}}
\newcommand{\eay}{\end{array}}
\def\RN{Reis\-sner-Nord\-str\"{o}m }
\def\rc{\rho_{\rm crit}}
\def\rl{\rho_\Lambda}
\def\rt{\rho_{\rm tot}}
\def\ie{{\it i.e.\;}}
\def\lp{\ell_{\rm Pl}}
\def\mp{m_{\rm Pl}}
\def\tp{t_{\rm Pl}}
\def\tf{t_{\rm FP}}
\def\om{\Omega_{\rm M}}
\def\oa{\Omega_{\Lambda}}
\def\ot{\Omega_{\rm tot}}
\def\tla{\widetilde{\lambda}_\ast}
\def\tom{\widetilde{\omega}_\ast}
\def\luv{\lambda_\ast^{\rm UV}}
\def\guv{g_\ast^{\rm UV}}
\def\lir{\lambda_\ast^{\rm IR}}
\def\gir{g_\ast^{\rm IR}}
\def\lir{\lambda_\ast^{\rm IR}}
\def\gir{g_\ast^{\rm IR}}

\addtolength{\oddsidemargin}{12mm}
\addtolength{\evensidemargin}{12mm}
\addtolength{\textwidth}{-24mm}
\addtolength{\topmargin}{11mm}
\addtolength{\footskip}{11mm}
\addtolength{\textheight}{-22mm}
\renewcommand{\baselinestretch}{1.5}
\textwidth16cm
\setlength{\oddsidemargin}{0cm}
\setlength{\jot}{0.3cm}                    

\begin{titlepage}
\renewcommand{\thefootnote}{\fnsymbol{footnote}}
\renewcommand{\baselinestretch}{1} 

\begin{flushright}
  MZ-TH/01-19, INFNCT/5/01 \\
\end{flushright}   

\begin{center}

{\Large \sc Cosmology with Self-Adjusting \\
[5mm]
Vacuum Energy Density from a \\
[6mm]
Renormalization Group Fixed Point}
\vspace{1cm}

{\large 
A. Bonanno} \\
\vspace{0.5cm}
\noindent
{\it Osservatorio Astrofisico,
Via S.Sofia 78, I-95123 Catania, Italy\\
INFN Sezione di Catania, Corso Italia 57, I-95129 Catania, Italy}

\vspace{1cm}
{\large
M. Reuter}\\
\vspace{0.5cm}
\noindent
{\it Institut f\"ur Physik, Universit\"at Mainz\\
Staudingerweg 7, D-55099 Mainz, Germany}
\end{center}

\begin{abstract}
Cosmologies with a time dependent Newton constant and cosmological constant are 
investigated. The scale dependence of $G$ and $\Lambda$ is governed by a set of renormalization
group equations which is coupled to Einstein's equation in a consistent way. The existence of 
an infrared attractive renormalization group fixed point is postulated, and the cosmological 
implications of this assumption are explored. It turns out that in the late Universe the
vacuum energy density is automatically adjusted so as to equal precisely the matter energy density, 
and that the deceleration parameter approaches $q = -1/4$. This scenario might explain the data
from recent observations of high redshift type Ia Supernovae and the cosmic microwave background 
radiation without introducing a quintessence field.
\end{abstract}

\end{titlepage}                                                 

\section{Introduction} 
Recent astronomical observations of high redshift type Ia Supernovae performed by two groups 
\cite{perl,riess,riess2} as well as the power spectrum of the cosmic microwave background 
radiation obtained by the BOOMERANG \cite{boom} and MAXIMA-1 \cite{max} 
experiments seem to indicate that 
at present the Universe is in a state of accelerated expansion. 
If one analyzes these data
within the 
Friedmann-Robertson-Walker (FRW) standard model of cosmology their most natural interpretation
is that the Universe is spatially flat and that the (baryonic plus dark) matter density $\rho$ 
is about one third of the critical density $\rc$. Most interestingly, the dominant contribution
to the energy density is provided by the cosmological constant $\Lambda$. The vacuum energy density
\be\label{1}
\rl\equiv \Lambda/(8\pi G)
\ee
is about twice as large as $\rho$, {\it i.e.} about two thirds of the critical density. With
$\om\equiv \rho/\rc$, $\oa\equiv \rl/\rc$ and $\ot\equiv \om+\oa$:
\be\label{2}
\om \approx 1/3, \;\;\;\;\;\; \oa \approx 2/3, \;\;\;\;\;\;\ot \approx 1 .
\ee
This implies that the deceleration parameter $q$ is approximately $-1/2$. While originally the cosmological
constant problem \cite{coscon} was related to the question why $\Lambda$ is so unnaturally small, 
the discovery of the important role played by $\rl$ has shifted the emphasis toward the ``coincidence
problem'', the question why $\rho$ and $\rl$ happen to be of the same order of magnitude precisely at this
very moment \cite{coscon2}.

In an attempt at resolving this naturalness problem the quintessence models \cite{quint,q2}
have been proposed in which the cosmological constant becomes a time dependent
quantity \cite{mrcw}. Their dynamics is arranged in such a way that $\rl$ automatically adjusts 
itself relative
to the value of $\rho$. In  most of the quintessence models 
the vacuum energy density $\rl$ is carried by a scalar field which has to be introduced 
on an ad hoc basis. 

In this paper we shall describe a different scenario which provides a natural explanation for 
the approximate equality of $\rho$ and $\rl$ and for the smallness of the cosmological constant. 
We are going
to set up a very general framework for cosmologies in which both Newton's constant and the cosmological
constant are time dependent. Within this framework, we shall formulate a single hypothesis whose 
consequences will be analyzed and which will turn out to imply that $\rho$ and $\rl$ are approximately
equal in the late Universe. 
In a nutshell, the hypothesis is that there exists an infrared (IR) 
attractive fixed point for the renormalization
group (RG) flow of the (dimensionless) Newton constant and cosmological constant, respectively. 
Our analysis 
is at purely phenomenological level in the sense that it is not necessary to know the details of the 
physics which is responsible for this postulated fixed point. 

Before we formulate our hypothesis in detail
we first describe the kinematic framework we are going to employ. It is the same framework which we used
in ref.\cite{uv}, henceforth referred to as 
(I), for an analysis of the quantum gravity effects in the early Universe (Planck era). 
\section{The Kinematic Framework}
We consider homogeneous and isotropic cosmologies described by a standard Robertson-Walker metric
containing the scale factor $a(t)$ and the parameter $K=0,\pm 1$ which distinguishes the three
types of maximally symmetric 3-spaces of constant cosmological time $t$. The dynamics is governed
by Einstein's equation $R_{\mu\nu}-\frac{1}{2}g_{\mu\nu}R = -\Lambda g_{\mu\nu}+
8\pi G T_{\mu\nu}$ with a conserved energy-momentum tensor ${T_{\mu}}^\nu={\rm diag}(-\rho,p,p,p)$
for which we assume the equation of state $p(t) = w \rho (t)$ where $w>-1$ is an arbitrary
constant. Now we perform a ``RG improvement'' \cite{mr,bh2,bh1} of Einstein's equation by replacing
$G\rightarrow G(t)$ and $\Lambda \rightarrow \Lambda (t)$ where the time dependence of $G$ and $\Lambda$
is such that the integrability of the field equations is maintained. This leads to the 
following system of equations:
\begin{subequations}
\ba
&&\Big ({\dot{a}\over a}\Big )^2+{K\over a^2}= 
{1\over 3}\Lambda+{8\pi\over 3} G\; \rho\label{3.a}\\[2mm]
&&\dot{\rho}+3(1+w)\; \Big ({\dot{a}\over a}\Big )\; \rho= 0\label{3.b}\\[2mm]
&&\dot{\Lambda}+8\pi\rho \; \dot{G} = 0\label{3.c}\\[2mm]
&&G(t)\equiv G(k=k(t)),\;\;\; \Lambda(t) \equiv \Lambda(k=k(t))\label{3.d}
\ea
\end{subequations}
Eq.(\ref{3.a}) is the standard Friedmann equation with a time dependent $\Lambda$ and $G$,
and Eq.(\ref{3.b}) expresses the conservation of $T^{\mu\nu}$. Eq.(\ref{3.c}) is a novel
consistency condition which is dictated by Bianchi's identity. It guarantees that the RHS
of Einstein's equation has vanishing covariant divergence. The system of equations 
(2.1a,b,c) has already been studied in the literature 
\cite{sys1,sys2}. Our crucial new ingredient \cite{uv} are the RG
equations (\ref{3.d}). Their meaning is as follows.

We describe gravitational phenomena at a typical distance scale $\ell\equiv k^{-1}$ in terms of a scale
dependent effective action $\Gamma_k[g_{\mu\nu}]$ which should be thought of as a Wilsonian
coarse grained free energy functional. The mass parameter $k$ is a IR cutoff in the sense that
$\Gamma_k$ encapsulates the effect of all metric fluctuations with momenta larger than $k$
while those with smaller momenta are not yet ``integrated out''. When evaluated at tree level, 
$\Gamma_k$ describes all processes involving a single characteristic momentum $k$ with all 
loop effects included. 

In \cite{mr}, $\Gamma_k$ has been identified with the effective average action \cite{berg} for
Euclidean quantum gravity and an exact functional RG equation for the $k$-dependence
of $\Gamma_k$ has been derived. Nonperturbative solutions were obtained within the 
``Einstein-Hilbert truncation'' which assumes $\Gamma_k$ to be of the form 
\be\label{4}
\Gamma_k=
(16\pi G(k))^{-1}\int d^4 x \sqrt{g} \{ -R(g) + 2\Lambda(k)\} 
\ee
The RG equations yield an explicit answer for the $k$-dependence of the running Newton constant
$G(k)$ and the running cosmological constant $\Lambda(k)$. Within the Einstein-Hilbert approximation,
the renormalization effects are strong only if $k$ is close to the Planck mass $\mp$. In (I) we argued
that they are important for an understanding of the Planck era immediately after the big bang. 
However, there are indications \cite{tsamis} that quantum Einstein gravity, because of its
inherent IR divergences, is subject to strong renormalization effects also at very large distances.
In cosmology those effects would be relevant to the Universe at late times. It has been speculated
that they might lead to a dynamical relaxation of $\Lambda$, thus solving the cosmological 
constant problem \cite{tsamis}. An analysis of such IR effects in the framework of the effective average
action is not available yet. It would require truncations which are much more complicated than
(\ref{4}) and which contain nonlocal invariants \cite{nonloc}, for instance. 

Nevertheless, in order to describe the idea of the ``RG improvement'' let us assume that we actually
know the functions $G(k)$ and $\Lambda(k)$ for all values of $k$, in particular for $k\rightarrow 0$,
{\it i.e.} in the IR. The idea is to express the mass parameter $k$ in terms of the physically 
relevant cutoff scale. In (I) we argued that, in leading order, the correct cutoff identification
in a Robertson-Walker spacetime is 
\be\label{5}
k(t) = \xi / t
\ee
where $\xi>0$ is an a priori unknown constant. Inserting (\ref{5}) into $G(k)$ and $\Lambda(k)$
we obtain the ${\it time}$ dependent quantities $G(t) \equiv G(k = \xi/t)$ and 
$\Lambda(t) \equiv \Lambda(k = \xi / t)$. This is precisely what is meant by the last two equations
of the system (2.1), Eq.(\ref{3.d}). (See (I) for further details.)

The cutoff identification (2.3) applies in the case of perfect homogeneity and isotropy 
for which $k_{\rm cosmo}\equiv k(t)=\xi/t$ is the only relevant scale. Allowing for 
(large, nonlinear) density perturbations $\delta \rho ({\bf x}, t)$ of a typical wave length 
$\lambda_{\rm pert}$ we introduce a new scale $k_{\rm pert}=2\pi / \lambda_{\rm pert}$
into the problem. Similarly, immersing a localized matter distribution (a massive body)
of total mass $M$ into the cosmological fluid gives rise to the scale $k_{\rm M}=M$. 
In the situations of interest the latter two mass scales are much larger than the
cosmological one: $k_{\rm pert}, k_{\rm M} \gg k_{\rm cosmo}$. 

In a situation with more than one possible physical cutoff scale the general theory of the
effective average action [18] implies that the relevant action is $\Gamma_k$ where
$k$ is the {\it largest} one of the various competing scales. Hence in order to describe
the physics of density perturbations of the localized matter distribution
one has to use $\Gamma_k$ at $k=k_{\rm pert}$ and $k=k_{\rm M}$, respectively,
rather than at $k=k_{\rm cosmo}$. In this case one needs to know $G(k)$ and 
$\Lambda(k)$ for $k$ near $k_{\rm pert}, k_{\rm M} \gg k_{\rm cosmo}$
{\it i.e.} in a very different regime. 

In the present paper we are interested only in the large scale dynamics of the Universe.
In the next section we are going to formulate a hypothesis on the running of 
$G(k)$, $\Lambda(k)$ for $k$ near $k_{\rm cosmo}$. The only assumption which we make about
the RG flow of $G$ and $\Lambda$ at ``non-cosmological'' scales $k\gg k_{\rm cosmo}$
(e.g. for $k\approx k_{\rm pert}, k_M$) is that at those scales the $k$-dependence
is very weak or zero so that standard gravity is recovered at sub-cosmological scales.

At this point we emphasize that it is not important for the present discussion which physical mechanism
actually causes the $k$- or $t$- dependence of $G$ and $\Lambda$. In particular, we also cover 
the possibility of an entirely {\it classical} origin of this running. In fact, it has been pointed out
\cite{classav,muk} that $G$ and $\Lambda$ naturally acquire a scale dependence if one starts
from a density distribution which is inhomogeneous at small distances and then 
performs spatial averaging over 3-volumes of increasing linear extension $\ell = k^{-1}$. The
classical dynamics of the averaged quantities leads to a nontrivial RG flow of $G$ and
$\Lambda$. Since the Universe is certainly not homogeneous at small distance scales, knowledge 
of this RG flow is important if one wants to parametrize observational data obtained at those
scales in terms of a homogeneous FRW model; it was argued that this classical scale dependence might
resolve the controversy about the value of the Hubble constant \cite{classav}. Another intriguing
result is that, after spatial averaging, the backreaction of long wavelength scalar and tensor
cosmological perturbations amounts to an effective negative energy density which counteracts
any pre-existing cosmological constant \cite{muk}.
\section{the fixed point hypothesis}
The above remarks complete our motivation of the system of equations (2.1). We shall now
formulate a hypothesis about the RG behavior of $G$ and $\Lambda$ whose dynamical origin is left open. 
Introducing dimensionless quantities $g(k)\equiv k^2 G(k)$ and $\lambda(k) \equiv \Lambda(k)/k^2$
the hypothesis is that for $k\rightarrow 0$ both $g$ and $\lambda$ run into an IR attractive 
non-Gaussian fixed
point, {\it i.e.} that for a wide range of initial conditions
\be\label{6}
\lim_{k\rightarrow 0} g(k) = \gir, \;\;\;\; \lim_{k\rightarrow 0}\lambda(k) = \lir
\ee
where $\gir$ and $\lir$ are strictly positive. While there are encouraging indications
pointing toward the existence of this fixed point \cite{tsamis}, a rigorous proof would be a formidable
task, however, probably comparable to a proof of confinement in QCD. In the following we explore
the cosmological implications of (\ref{6}) which, as we shall see, provide further evidence
for the fixed point hypothesis from the phenomenological side.

The postulated fixed point is the IR 
counterpart of the UV attractive non-Gaussian 
fixed point which is known to exist in the Einstein-Hilbert
truncation of pure quantum gravity \cite{souma,mr,ol}. For a large class
of trajectories \cite{frank},
\be\label{7}
\lim_{k\rightarrow \infty} g(k) = \guv, \;\;\;\; 
\lim_{k\rightarrow\infty}\lambda(k) = \luv
\ee
More generally, we assume that the exact cosmologically relevant RG trajectory in $(g,\lambda)$-space
smoothly interpolates between $(\guv,\luv)$ for $k\rightarrow \infty$ and $(\gir,\lir)$ for 
$k\rightarrow 0$. The UV fixed point is important for the very early Universe 
$(t\rightarrow 0)$ while the IR fixed point determines the cosmology at late times $(t\rightarrow \infty)$.

It is reassuring to note that a similar crossover between two non-trivial 
RG fixed points has actually 
been shown to exist in 2-dimensional Liouville quantum gravity \cite{liouv}. Its RG trajectory
connects two conformal field theories with central charges $25-c$ and $26-c$, respectively, 
where $c$ is the central charge of the matter system.

In the vicinity of either of the two fixed points the evolution of the dimensionful $G$ and 
$\Lambda$ is approximately given by
\be\label{8}
G(k) = \frac{g_\ast}{k^2},\;\;\;\; \Lambda(k) = \lambda_\ast \; k^2
\ee
Here and in the following the fixed point values are denoted $g_\ast$ and $\lambda_\ast$ if the 
corresponding formula is valid both at the UV and at the IR fixed point. From
(\ref{8}) with (\ref{5}) we obtain the time dependent Newton constant and cosmological
constant:
\be\label{9}
G(t) = g_\ast \xi^{-2} \; t^{2},\;\;\;\; \Lambda(t) = \frac{\lambda_\ast\xi^2}{t^2}
\ee
The power laws (\ref{9}) are valid both for $t\searrow 0$ and, with different coefficients, 
for $t\rightarrow \infty$. By assumption, the time dependence of $G$ and 
$\Lambda$ at intermediate times is given by smooth functions $G(t)$ and $\Lambda(t)$ which 
interpolate between the UV and IR power laws. If we use these functions $G(t)$ and 
$\Lambda(t)$ in the coupled system (2.1), its solution gives us the scale factor 
$a(t)$ and the density $\rho (t)$ of the ``RG improved cosmology''. 

To be precise, our hypothesis about the RG trajectory $k\mapsto (G(k),\Lambda(k))$
consists of two parts. For  $k\lesssim k_{\rm cosmo}$ we assume the validity of the 
IR fixed point behavior (3.3). For  $k\gtrsim k_{\rm cosmo}$ the assumption 
is that $G$ and $\Lambda$ depend on $k$ only extremely weakly or are $k$-independent.
In this manner we recover standard gravity with $G,\Lambda$=const at the length scales
smaller than the cosmological scale $\propto t$. In particular $G$ and $\Lambda$
are essentially constant at $k_{\rm pert}$ and $k_{\rm M}$ so that the dynamics of 
localized matter distribution remains unchanged and there is no conflict with the 
classical tests of general relativity. Stated differently, the hypothesis is that the 
nontrivial runnning is due to quantum fluctuations with momenta between $k_{\rm cosmo}$
and $k_{\rm max}$ where $1/k_{\rm max}$ is the length scale characterizing the largest 
localized structures in the Universe of which we know for sure that standard gravity 
applies.

\section{cosmological solutions in the fixed point regime}
It is important to note that the system (2.1) is actually overdetermined: it consists of 
5 equations for the 4 unknowns $a,\rho,G$ and $\Lambda$. This leads to nontrivial 
consistency conditions for admissible RG trajectories and cutoff identifications. 
(See (I) for a detailed discussion.)
In (I) we showed that if $G(t)$ and $\Lambda(t)$ are given by (\ref{9}) the system (2.1)
has indeed a consistent solution provided $\xi$ assumes a specific value. Quite generally the
consistency conditions have the very welcome feature of fixing the ambiguities in the modeling
of the cutoff (here $\xi$) to some extent.

For the case of a spatially flat Universe $(K=0)$ the consistency condition reads
\be\label{10}
\xi^2 =\frac{8}{3(1+w)^2\;\lambda_\ast}
\ee
If it is satisfied, the system (2.1) with (\ref{9}) has the following almost unique
solution
\begin{subequations}
\ba\label{11a}
&&a(t) = \Big [ \Big ({3\over 8}\Big )^2 (1+w)^4 \; g_\ast \lambda_\ast \; {\cal M} \Big ]^{1/(3+3w)}\; t^{4/(3+3w)}\\[2mm]
&&\rho(t) = {8\over 9\pi (1+w)^4 \; g_\ast \lambda_\ast}\; {1\over t^4}\label{11b}\\[2mm]
&&G(t) = {3\over 8}\;(1+w)^2 \; g_\ast \lambda_\ast \; t^2\label{11c}\\[2mm]
&&\Lambda(t) = {8\over 3(1+w)^2}\; {1\over t^2}\label{11d}
\ea
\end{subequations}                                                            
Apart from the parameter $w$ and the product $g_\ast \lambda_\ast$, the solution (4.2) depends
only on a single constant of integration, ${\cal M}$, whose value affects only the
overall scale of $a(t)$. Numerically it equals $8\pi \rho(t) [a(t)]^{3+3w}\equiv {\cal M}$
which, like in standard cosmology, is a conserved quantity for the system (2.1).

The solution (4.2) has several very interesting and attractive features. Introducing the 
critical density 
\be\label{12}
\rc (t) \equiv \frac{3}{8\pi G(t)}\;\Big ( \frac{\dot{a}}{a} \Big )^2
\ee
we find for any value of $w$, $g_\ast\lambda_\ast$, and ${\cal M}$ that 
$\rc (t) = 2 \rho (t)$ and $\rho_\Lambda (t) = \rho (t)$. Hence 
\be\label{13}
\rho=\rho_\Lambda
=\frac{1}{2}\rc
\ee
Thus the total energy density $\rt \equiv \rho +\rl$ equals precisely the critical one:
$\rt (t) = \rc (t)$. This latter equality does not come as a surprise because also
the RG improved Friedmann equation can be brought to the form
\be\label{14}
K = \dot{a}^2 \; [ \rt/\rc -1]
\ee
so that $\rt = \rc$ holds true for any solution with $K=0$. On the other hand, the
exact equality of the matter energy density $\rho$ and the vacuum energy density 
$\rl$ is a nontrivial prediction of the fixed point solution. In terms of the
relative densities, 
\be\label{15}
\om = \oa = \frac{1}{2}, \;\;\;\;\;\;\;\;\; \ot =1
\ee
Also the Hubble parameter of the solution (4.2)
\be\label{16}
H \equiv \frac{\dot{a}}{a} = \frac{4}{3+3w}\; \frac{1}{t}
\ee
and its deceleration parameter
\be\label{17}
q \equiv -\frac{a\,\ddot{a}}{\dot{a}^2} = \frac{3w-1}{4}
\ee
are independent of $g_\ast$, $\lambda_\ast$ and ${\cal M}$. It can be shown
that the standard formula for $q$ in terms of the relative densities
continues to be correct for the improved system (2.1) with an arbitrary RG
solution (2.1d):
\be\label{18}
q = \frac{1}{2} \; (3w+1)\;\om-\oa
\ee
Clearly (\ref{18}) is satisfied by (\ref{17}) with (\ref{15}).

Another interesting feature of the fixed point solution is that it yields a universal,
time independent value of the ``Machian'' quantity $\rho \; G \; t^2$ \cite{mach}
\be\label{19}
\rho (t) \; G(t) \; t^2 = \frac{1}{3\pi (1+w)^2}
\ee

Since the 1-parameter family of cosmologies (with parameter ${\cal M}$ ) described by 
(4.2) is the most general solution of the system (2.1) with the fixed point running (\ref{9}) we
conclude that {\it every} complete solution of (2.1), valid for all $t\in (0,\infty)$, 
approaches one of the solutions (4.2) either for $t\searrow 0$ or for $t\rightarrow \infty$,
depending on whether the fixed point is UV or IR for the trajectory considered. Thus
the fixed point solution (4.2) is an attractor in the space of the functions
$(a,\rho,G,\Lambda)$.

Note that the product $G(t)\Lambda(t)= G(k)\Lambda(k) =g_\ast \lambda_\ast$ is
constant in the vicinity of any fixed point. Its actual value is
characteristic of this fixed point. While, for pure gravity,
$\guv \luv = O(1)$ at the UV fixed point of [12], 
the hypothetical IR fixed point of the coupled gravity-matter system
has $\gir \lir = O(10^{-120})$. It is important to understand that the smallness of
this number does not pose any finetuning problem as in the standard situation. In fact,
in our approach both $g_\ast^{\rm UV}$ $\lambda_\ast^{\rm UV}$
and $g_\ast^{\rm IR}$ $\lambda_\ast^{\rm IR}$ are fixed and 
well-defined numbers which, at least in principle, can be computed
from the RG equation. However, apart from being a difficult task 
technically, their actual determination is possible
only once we know the complete system of all matter fields in the Universe. The number
$10^{-120}$ reflects specific properties of this matter system coupled to gravity rather than an
initial condition. 

In the following we focus on the possibility of an IR attractive fixed point which governs
the cosmological evolution for $t\rightarrow \infty$. Whether or not there exists in addition
an UV fixed point approached for $t\searrow 0$ is not important in this context.
Assuming the existence of the IR fixed point and the validity of the equations (2.1) we are led 
to conclude that the late Universe, for which the RG trajectory is already sufficiently
close to the fixed point, is described by the power laws (4.2). This leads to the unambiguous
prediction that $\om = \oa = {1}/{2}$ for every value of $w$. Moreover,
if we make the additional assumption that the late Universe is matter dominated
($w=0$), Eq.(\ref{17}) yields $a\propto t^{4/3}$ with the deceleration parameter
$-1/4$. Hence, near the fixed point, 
\be\label{20}
\om = \oa = \frac{1}{2}, \;\;\;\;\;\;\;\; q = -\frac{1}{4} \;\;\;\;\;\;\;\;\; (w=0)
\ee
Without any further input from the RG equations (which would require detailed knowledge of 
the matter sector) we cannot assess at which time $\tf$ the fixed point behavior sets in.
At $\tf$ a generic solution, arising from arbitrary initial conditions, starts looking like (4.2).
However, it is very intriguing that the prediction (\ref{20}), valid for $t>\tf$, is quite 
close to the values (\ref{2}) favored by the recent observations\footnote{Note that the values (\ref{2})
are still afflicted with large error bars \cite{perl}. In the 
$(\om,\oa)$-plane, the values (\ref{20}) lie within the ellipse corresponding to the $2\sigma$ 
confidence region.}. In particular, the fixed point structure provides a natural
explanation for the mysterious equality (or approximate equality) of $\rho$ and
$\rl$. This success supports the idea that {\it the present-day Universe 
is in the, or at least close to the IR fixed point regime.}

The deviation of the observed values (\ref{2}) from $\om=\oa=1/2$ could be due to the fact
that the fixed point behavior is not fully developed yet so that the Universe still has
some way to go before the finer quantitative details of the solution (4.2) are realized.
However, given the large observational uncertainties \cite{perl} it is also well possible
that more precise observations will lead to modified values of $\om$ and $\oa$ which are
closer to $\om=\oa=1/2$. 

A further testable prediction of the fixed point hypothesis is the time variation of 
Newton's constant. From (4.2) we obtain 
\be\label{22}
\frac{\dot{G}}{G}=\frac{2}{t}=\frac{1}{2}\;(3+3 w) \; H(t)
\ee
The experimental upper bound from laboratory and Solar System experiments for the present-day
value of this quantity \cite{gill} is of the order of $|\dot{G}/G|\lesssim 
(10^{11} {\rm yr})^{-1}$.
Hence even the technology available today is not very far away from being able to verify
or falsify (\ref{22}). One should bear in mind, however, that the $G$ in Eq.(\ref{22}) refers to a 
different length scale than the one measured in Solar System experiments, say.

In this context it is interesting to remark that recently a Brans-Dicke theory with a quadratic
self-coupling of the Brans-Dicke field has been constructed \cite{berto} which admits
a solution very similar to our fixed point solution (4.2) and which predicts the same time
dependence of Newton's constant.

Up to now we discussed the spatially flat Universe only. The $K=0$-solution (4.2) exists
for every value of the parameter $w$. The situation is different when we now look for
solutions with $K=\pm 1$ corresponding to spatially curved Universes. In (I) we have shown that
consistent solutions to the system (2.1), (\ref{9}) with $K= + 1$ or $K=-1$ exist only if 
$w=+1/3$,  {\it i.e.} for a radiation dominated Universe. Assuming the validity of (2.1) and of 
the fixed point hypothesis, and excluding the possibility of $w=+1/3$ for $t\rightarrow \infty$,
we see that the Universe can fall into the basin of attraction induced by the IR
fixed point only if it is spatially flat, {\it i.e.} if $K=0$. By Eq.(\ref{14}) this
is equivalent to $\rt = \rc$, as it would be in standard cosmology.

It is important to stress that the fixed-point scenario sets in at scales that are 
much larger than the characteristic scales of local gravitational interactions.
One cannot expect that any simple scaling law, or cutoff identification, can 
be used in a general gravitating system when we are probably far from any
fixed point in the $G$ and $\Lambda$ evolution. Instead we assume that 
$G$ and $\Lambda$ can be RG evolved only through the large scale evolution
of the Universe, which would then provide a natural scaling law and a 
meaningful cutoff identification by means of  the cosmological time.
In fact at much smaller scales than galaxy cluster scales
neither we can establish the flow equations, nor we could guess a meaningful identification
of the cutoff since several ``crossover'' regions can be present, which would drive us away from the 
fixed point law and would make any identification of the cutoff problematic.

It is nevertheless possible to describe the evolution of weak, 
localized density perturbations $\delta\rho \ll \rho$ within our framework in 
a consistent way. The reason is that, at the linearized level, the effective
cutoff scale is still given by $k_{\rm cosmo}$ rather than the scales 
associates with $\delta \rho$.

Let us consider a ``fundamental observer'' describing the cosmological fluid
flow lines. Its 4-velocity is
$u^\mu= dx^\mu/ d\tau , \;\;\; u^\mu u_\mu = -1$
where $\tau$ is the proper time along the fluid flow lines.
The projection tensor in the tangent 3-space orthogonal to $u^\mu$
is $h_{\mu\nu} = g_{\mu\nu}+u_\mu u_\nu$ with  
${h^\mu}_\nu {h^{\nu}}_\sigma = 
{h^\mu}_\sigma$ and ${h^{\mu}}_\nu u^\nu =0$. The energy-momentum tensor then reads
$T^{\mu\nu} = \rho \; u^\mu u^\nu + p\;h^{\mu\nu}$, where $\rho = \bar{\rho}+
\delta \rho(x^\mu)$ and $p=\bar{p}+\delta p(x^\mu)$ being $\bar{\rho}$ and
$\bar{p}$ the pressure and the density of the unperturbed universe.
From the Bianchi identities and the conservation law $\nabla_\nu{T^{\mu\nu}}=0$   
we have the following propagation equations
\ba
&& u^\nu\; (\nabla_\nu\Lambda + 8\pi \rho \;\nabla_\nu G)=0\\[2mm]
&& h^{\mu\nu}\;( -\nabla_\nu{\Lambda} + 8\pi p \;\nabla_\nu G) = 0
\ea
for the $u^\mu$ direction and for the tangent orthogonal 3-space, respectively, together with 
\be\label{lg}
G\Lambda = \lambda_\ast g_\ast
\ee
coming from the fixed point behavior (\ref{8}). From (4.14) we thus see that in the late Universe, which 
is of interest for us, when $\bar{p}=0$ is a good equation of state, there are no space gradients 
of the cosmological constant up to first order perturbation theory. In fact since 
$\delta p$, $\delta\rho$ and the space gradients of $G$ and $\Lambda$ are assumed to be of the
same order, $\delta p\; {h^\nu}_\mu\nabla_\nu G$ is of second order. 
Therefore, by differentiation of (\ref{lg}) and subsequent projection onto 
the 3-space, one concludes that also the space gradients of $G$ are negligible in
this approximation. This result implies that the description of local gravitational interactions
depends only on the global, large scale, 
time evolution of $G$ and $\Lambda$ and it does not introduce additional 
effects coming from the space gradients of $G$ and $\Lambda$ that could 
in principle appear in the description of local deviations from homogeneity.
One can then discuss the standard scenario of structure formation
with the large scale structure evolution provided by the solution (4.2). 

We also emphasize that the standard experimental value of Newton's constant,
$G_{\rm exp}$, does not coincide with the value $G(k=\xi/t_0)$ which is relevant for
cosmology today, {\it i.e.} for $t=t_0$. $G_{\rm exp}$ is measured (today) at 
$k_{\rm exp}\propto \ell^{-1}$ where the length $\ell \equiv \ell_{\rm sol}$ is a 
typical solar system length scale, say. Thus, in terms of the running Newton constant,
$G_{\rm exp} = G(k=\xi'/\ell_{\rm sol})$,
since $\ell_{\rm sol}\ll t_0$, and since in presence of several scales the relevant
cutoff is always the larger one\footnote{See ref.[14] for a detailed discussion of this point.}.
It is only the cosmological quantity $G(k=\xi/t)$ which grows $\propto t^2$ in the fixed
point regime, not $G_{\rm exp}$.  
This remark entails that a $t^2$-growth of the cosmological Newton constant in the 
recent past does not ruin the predictions about primordial nucleosynthesis which requires that
$G(k=\xi/t_{\rm nucl})$ coincides with $G_{\rm exp}$ rather precisely. 
In fact, at the time $t=t_{\rm nucl}$ of nucleosynthesis the cosmological Newton 
constant was indeed $G(k=\xi/t_{\rm nucl})\approx G_{\rm exp}$ 
since $ct_{\rm nucl}$ and $\ell_{\rm sol}$
are of the same order of magnitude (a few light minutes).
\section{discussion and conclusion}
In this paper we modified Einstein's equation for a Robertson-Walker spacetime by allowing
for a scale-, and hence time-dependent Newton's constant and cosmological constant.
The scale dependence of $G$ and $\Lambda$ follows from a renormalization group which 
could be of either classical or quantum origin. We postulated that the RG flow at large
distances is governed by an IR fixed point and we investigated the cosmological
implications of this assumption. It turned out that, in the fixed point regime, 
the vacuum energy density $\rl$ equals precisely the matter density $\rho$ and that they decrease
proportional to $1/t^4$, while Newton's constant increases $\propto t^2$. Assuming that the present
Universe is in that regime, this scenario leads to a natural resolution of the
coincidence problem (``Why is $\rho/\rl=O(1)$ today?'') and of the cosmological constant 
problem in its original form (``Why is $\Lambda$ so small?''). It predicts that
the universe is spatially flat.

Obviously cosmologies of the type found here are very attractive from the phenomenological
point of view. This success provides a strong motivation for further attempts at actually
proving the existence of the postulated IR fixed point and the validity of the improved
system of cosmological evolution equations, Eqs.(2.1).

In the present paper we assumed the existence of a cosmologically relevant IR fixed point, while in 
(I) we investigated the consequences of a UV fixed point for the Planck era directly after
the big bang. The UV fixed point has been shown to exist in (the Einstein-Hilbert truncation of) 
pure quantum gravity. The assumption that the matter contents of the Universe is such that there
exists both the UV and the IR fixed point leads to a particularly symmetric cosmological scenario:
The Universe begins and ends at two different attractors, attractive for $t\searrow 0$ and
$t\rightarrow \infty$, respectively, and its evolution between them is a kind of crossover between
two fixed points.

Before closing let us see what happens if we relax our hypothesis to some extent. Up to now
we assumed the IR fixed point to be attractive in all directions in the space of coupling constants.
It might be that actually there are also unstable directions so that
for $t\rightarrow \infty$ the Universe is eventually driven away from it.
Nevertheless, if it stays for a sufficiently long time close to the fixed point, the solution (4.2)
might still be a rather accurate description of the present Universe even though its
ultimate fate for $t\rightarrow \infty$ cannot be predicted then.

As a more radical step, let us give up our idea that the $t$-dependence of $G(t)$ and
$\Lambda(t)$ arises from some RG trajectory $G(k)$ and $\Lambda(k)$ by identifying $k\equiv k(t)$.
We retain however the three differential equations (\ref{3.a}), (\ref{3.b}), and (\ref{3.c}).
The system (2.1a,b,c) without (\ref{3.d}) is underdetermined so that an additional condition on 
$a,\rho,G$ and $\Lambda$ may be imposed. Without providing a physical explanation, 
it was assumed in ref.\cite{sys2} 
that Newton's constant varies according to a power law
$G\propto t^n$ with an arbitrary, not necessarily integer exponent $n$. With this additional 
condition the system (2.1a,b,c) with $K=0$ has the following 2-parameter family of solutions
(The parameters are ${\cal M}$ and $C>0$.) : 
\begin{subequations}
\ba\label{23a}
&&a(t) = \Big [\frac{3(1+w)^2}{2(n+2)}\;{\cal M}\; C\Big ]^{1/(3+3w)}\;t^{(n+2)/(3+3w)}\\[2mm]
&&\rho(t) = \frac{(n+2)}{12\pi\;(1+w)^2\;C}\;\frac{1}{t^{n+2}}\label{23b}\\[2mm]
&&G(t) = C\; t^n\label{23c}\\[2mm]
&&\Lambda(t) = \frac{n(n+2)}{3(1+w)^2}\;\frac{1}{t^2}\label{23d}
\ea
\end{subequations}
The solution (4.2) resulting from the fixed point hypothesis corresponds to the special case 
$n=2$ and $C=3(1+w)^2 g_\ast\lambda_\ast/8$. Note that the exponent $n=2$ is an unambiguous
prediction of the fixed point scenario. It is obtained even if we use an identification of 
$k$ in terms of $t$ which is different from $k=\xi/t$. The reason is that (\ref{8}) 
implies $G\Lambda=g_\ast \lambda_\ast =\text{const}$ for every function $k=k(t)$, but
according to (5.1) the product $G\Lambda$ is constant only if $n=2$. For the cosmology
(5.1) one easily computes
\be\label{24}
\om= \frac{2}{n+2},\;\;\;\oa={n\over n+2}, \;\;\;\; \ot =1, \;\;\;\; q = {1+3w-n\over n+2}
\ee
For $n\not =2$, $\rho$ and $\rl$ are no longer equal: $\oa/\om= n/2$. It is amusing to note that 
setting $n=4$ and $w=0$ yields 
\be\label{25}
\om = \frac{1}{3}, \;\;\;\;\; \oa = \frac{2}{3}, \;\;\;\;\; q = -{1\over 2}\;\;\;\;\;\; (n=4,w=0)
\ee
which equals quite precisely the values (1.2) favored by the present experimental data. 

Once more we see that a more accurate experimental determination of $\om$ and $\oa$ 
is highly desirable. In case the values ultimately stabilize near $\om=\oa=1/2$
this would be an important step toward confirming the fixed point scenario. 
If instead they remain close to their present values (\ref{2}) it might be 
worthwhile to reconsider the more general cosmologies of the 
type (5.1) with $n=4$. However, at present there seems to be no theoretical 
argument which would single out $G\propto t^n$ and $n=4$. While the classical
cosmological tests related to the early Universe 
(nucleosynthesis) are most probably insensitive to the modifications caused by the IR
fixed point, measurements of $\dot{G}/G$ at different length scales are another important test for the
validity of the fixed point hypothesis.
\section*{Acknowledgements}
M.R. would like to thank the Department of Theoretical Physics, University of Catania, the
Department of Theoretical Physics, University of Trieste, and the Astrophysical 
Observatory of Catania for their hospitality while this work was in progress. He also acknowledges
the financial support by INFN, MURST, and by a NATO traveling grant. 

\end{document}